\documentclass[8pt,letterpaper,twocolumn]{extarticle}

\usepackage[margin=1in]{geometry}
\usepackage{setspace}
\usepackage{parskip}             % blank line between paragraphs, no indent
\usepackage[utf8]{inputenc}
\usepackage[T1]{fontenc}
\usepackage{newtxtext,newtxmath}

\usepackage{graphicx}
\usepackage{booktabs}
\usepackage{array}
\usepackage{tabularx}
\usepackage{xcolor}
\usepackage{listings}
\usepackage{hyperref}
\usepackage{enumitem}
\usepackage{titlesec}
\usepackage{caption}
\usepackage{dblfloatfix}

\hypersetup{
    colorlinks=true,
    linkcolor=black,
    citecolor=blue!50!black,
    urlcolor=blue!60!black
}

\titleformat{\section}{\large\bfseries}{\thesection}{0.6em}{}
\titleformat{\subsection}{\normalsize\bfseries}{\thesubsection}{0.5em}{}
\titlespacing*{\section}{0pt}{1.2em}{0.4em}
\titlespacing*{\subsection}{0pt}{0.8em}{0.3em}

\definecolor{codebg}{RGB}{248,248,248}
\definecolor{codekw}{RGB}{0,70,140}
\definecolor{codecmt}{RGB}{120,120,120}
\definecolor{codestr}{RGB}{0,120,60}
\lstdefinestyle{code}{
  backgroundcolor=\color{codebg},
  basicstyle=\ttfamily\footnotesize,
  keywordstyle=\color{codekw}\bfseries,
  commentstyle=\color{codecmt}\itshape,
  stringstyle=\color{codestr},
  breaklines=true,
  showstringspaces=false,
  frame=single,
  rulecolor=\color{gray!40},
  language=Python,
  aboveskip=4pt,
  belowskip=4pt
}
\title{\Huge \textbf{CommitLLM:} A Fine-Tuned Pipeline for \\ Git Commit Message Generation}
\author{
  Md Rafid Haque \\
  \small University of Illinois at Chicago \\
  \small mhaqu23@uic.edu
  \and
  Poojan Narendrabhai Patel \\
  \small University of Illinois at Chicago \\
  \small ppate539@uic.edu
  \and
  Meetkumar Vijaybhai Raychura \\
  \small University of Illinois at Chicago \\
  \small mrayc@uic.edu
  \\[0.8em]
}
\date{}

\begin{document}
\maketitle

% =====================================================================
\begin{abstract}
Developers frequently write uninformative git commit messages such as \texttt{"fix"} or \texttt{"update stuff"}, degrading the value of version-control history for code review, debugging, and onboarding. We present \textbf{CommitLLM}, a three-stage pipeline that generates concise, Conventional Commits-compliant messages from code diffs using a fine-tuned small language model. The system combines (1)~QLoRA fine-tuning of Mistral-7B-Instruct-v0.2 on the CommitPackFT dataset, (2)~constrained decoding to enforce brevity, and (3)~deterministic post-processing to strip conversational artifacts and enforce format. On a 50-sample evaluation, CommitLLM achieves 98\% format compliance (vs.\ 22\% for vanilla Mistral), reduces average output length from 154.8 to 37.9 characters, and improves LLM-as-a-Judge scores from 1.97 to 3.68 out of 5. Notably, the post-processing layers contribute more to quality improvement than the fine-tuning itself, suggesting that for structured-output tasks, treating the LLM as a component in a deterministic pipeline is more effective than optimizing the model alone. The entire system runs on a single consumer GPU (NVIDIA T4, 16~GB VRAM).
\end{abstract}

% =====================================================================
\section{Introduction}

\subsection{Motivation}

Most developers either skip writing meaningful commit messages or write things like \texttt{"fix"} or \texttt{"update stuff"}. A model that reads a diff and produces a clean, Conventional Commits-style message would save time and make commit history actually useful for code review and debugging. This paper asks a practical question: can a small open-source LLM be fine-tuned to write good git commit messages from code diffs, on hardware that any developer has access to?

\subsection{Goals}

Our goal was twofold. First, fine-tune Mistral-7B-Instruct-v0.2 on a diff-to-commit dataset using parameter-efficient methods so it could run on a consumer GPU (Google Colab T4). Second, build a complete inference pipeline that reliably produces short, well-formatted commit messages, not just a model that sometimes gets it right.

We originally planned to package the model as a VS Code extension, but the engineering effort of local model deployment (shipping weights, hosting endpoints, or assuming a local inference server) proved orthogonal to the core research question. We instead invested in building a more rigorous evaluation harness and discuss the IDE integration as future work (Section~\ref{sec:future}).

\subsection{Assumptions}

We made a few assumptions and stuck with them throughout:

\begin{itemize}[leftmargin=1.5em,itemsep=0.2em]
  \item \textbf{Compute:} Google Colab with a single NVIDIA T4 GPU (16~GB VRAM). Everything we did had to fit in this budget.
  \item \textbf{Language:} Python 3 with the HuggingFace ecosystem (\texttt{transformers}, \texttt{peft}, \texttt{trl}, \texttt{datasets}, \texttt{bitsandbytes}, \texttt{evaluate}).
  \item \textbf{Quantization assumption:} 4-bit NF4 quantization with bf16 compute would not significantly hurt model quality. The QLoRA paper~\cite{dettmers2023qlora} suggests this is a safe assumption and our results are consistent with that.
  \item \textbf{Dataset assumption:} CommitPackFT~\cite{commitpackft} is good enough as-is. We did not try to clean it further beyond what BigCode already did.
\end{itemize}

\subsection{Definitions}
\label{sec:defs}

A few terms come up enough in this paper that we should pin them down:

\begin{itemize}[leftmargin=1.5em,itemsep=0.2em]
  \item \textbf{LoRA (Low-Rank Adaptation)~\cite{hu2021lora}:} a parameter-efficient fine-tuning method that freezes the base model and trains small rank-$r$ matrices inserted next to the original weights.
  \item \textbf{QLoRA~\cite{dettmers2023qlora}:} LoRA on top of a 4-bit quantized base model. This is what makes 7B fine-tuning fit on a T4.
  \item \textbf{NF4:} a 4-bit quantization data type designed to be optimal for normally-distributed weights, used by QLoRA.
  \item \textbf{Conventional Commits:} an industry convention where commit messages start with a type prefix like \texttt{feat:}, \texttt{fix:}, or \texttt{docs:}~\cite{convcommits}.
  \item \textbf{BERTScore~\cite{zhang2020bertscore}:} a metric for text similarity that compares contextual embeddings rather than exact word overlap, so it understands paraphrases.
  \item \textbf{LLM-as-a-Judge~\cite{zheng2023judging}:} using a strong LLM to grade outputs from another model on a fixed rubric. We used Llama-3.3-70B and Llama-3.1-8B served via Groq.
\end{itemize}

\subsection{Summary of Approach}

We did three things, in this order:

\textbf{(1) Fine-tune.} We loaded Mistral-7B-Instruct-v0.2 in 4-bit NF4 and applied LoRA (rank 16, $\alpha=32$) on attention and MLP projection layers. We trained for one epoch on the CommitPackFT data using \texttt{SFTTrainer} from HuggingFace's TRL library. Total trainable parameters: about 42M out of 3.79B, or roughly 1.1\% of the model.

\textbf{(2) Build an inference pipeline, not just use the model.} The fine-tuned model on its own was still too verbose, it would sometimes produce paragraph-length output. So we wrapped it in two extra stages: aggressive constrained decoding (\texttt{max\_new\_tokens=20}, \texttt{temperature=0.1}, repetition penalty 1.1) and deterministic post-processing (slice after \texttt{[/INST]}, take first line only, strip conversational filler before the first colon if it is not a valid Conventional Commits type).

\textbf{(3) Evaluate.} We compared three systems, vanilla Mistral, raw fine-tuned Mistral, and the full pipeline, on four metrics: format compliance (regex), commit length, BERTScore semantic similarity to the human ground-truth, and LLM-as-a-Judge scores from Llama 3.3-70B / 3.1-8B.

% =====================================================================
\section{Background and Related Work}

\subsection{Tools and Libraries}

A few things we tried that did and did not work:

\begin{itemize}[leftmargin=1.5em,itemsep=0.2em]
  \item \textbf{HuggingFace TRL \texttt{SFTTrainer}:} worked out of the box once we figured out the new \texttt{SFTConfig} arguments. The \texttt{group\_by\_length=True} setting alone cut training time noticeably by reducing wasted padding.
  \item \textbf{\texttt{bitsandbytes} 4-bit quantization:} initially failed at inference time with a CUDA-related error; we had to upgrade to \texttt{bitsandbytes>=0.46.1} to make it work with the Colab T4.
  \item \textbf{Groq for LLM-as-a-Judge:} fast, cheap, generous free tier, but the daily token-per-day limit (100k tokens) forced a model swap from \texttt{llama-3.3-70b-versatile} to \texttt{llama-3.1-8b-instant} for the cleaned-output evaluation.
\end{itemize}

\subsection{Related Work}

For the modeling side, the two papers that mattered most were the original LoRA paper~\cite{hu2021lora} for the rank/$\alpha$/target-module choices, and the QLoRA paper~\cite{dettmers2023qlora} for the 4-bit NF4 + double-quantization recipe. The Mistral 7B technical report~\cite{jiang2023mistral} was useful for understanding why this base model is a good choice for fine-tuning at this scale. For the evaluation side, the BERTScore paper~\cite{zhang2020bertscore} explained why we should expect contextual-embedding similarity to be more forgiving than BLEU/ROUGE for short, paraphrase-heavy outputs like commit messages, and the MT-Bench / LLM-as-a-Judge paper~\cite{zheng2023judging} gave us the framing for using an LLM to grade outputs against a fixed rubric.

Key documentation resources included:

\begin{itemize}[leftmargin=1.5em,itemsep=0.2em]
  \item \textbf{HuggingFace PEFT documentation}~\cite{hfpeft} for the LoRA configuration API and target module selection.
  \item \textbf{HuggingFace TRL documentation}~\cite{hftrl} for \texttt{SFTTrainer} and \texttt{SFTConfig}.
  \item \textbf{The CommitPackFT dataset card}~\cite{commitpackft} for the data structure and licensing.
  \item \textbf{The Conventional Commits specification}~\cite{convcommits} for what ``compliant'' actually means as a target format.
  \item \textbf{The \texttt{bitsandbytes} documentation}~\cite{bnbdocs} for the right \texttt{BitsAndBytesConfig} settings.
  \item \textbf{Groq API docs}~\cite{groqdocs} for endpoint names, rate limits, and the async client.
\end{itemize}

The HuggingFace PEFT and TRL docs were the single most practically useful resources. The QLoRA paper was the most useful academic reference because it specifies exactly which 4-bit format to use and why. The Conventional Commits spec was unexpectedly important on the evaluation side: it gave us a concrete, externally-defined target for what a ``good'' commit looks like, which is what made the regex-based compliance metric possible.

% =====================================================================
\section{Approach}

\subsection{The Layered Strategy}

Our overall strategy can be summarized in one sentence: \emph{treat CommitLLM as a system, not as a model.} Most discussion of LLM fine-tuning focuses on getting better weights. We started there too, but quickly realized that even after a clean fine-tune, the model would still occasionally produce verbose, conversational, or multi-line output. For a tool that is supposed to drop into a developer's workflow, that is unacceptable, a commit-message generator that returns a paragraph 30\% of the time will get disabled within a day. So we structured the work around three layers, each one constraining the previous one further.

\textbf{Layer 1: The fine-tuned model.} The base model is Mistral-7B-Instruct-v0.2. We chose it because it is a 7B parameter model with strong instruction-following (it already understands the \texttt{[INST] ... [/INST]} chat format), it has a permissive license, and it is the right size for a T4 once quantized. Fine-tuning is QLoRA: 4-bit NF4 quantization on the base, LoRA adapters with rank 16 and $\alpha=32$ on attention ($q$, $k$, $v$, $o$) and MLP ($gate$, $up$, $down$) projections. One epoch over the CommitPackFT data, AdamW (paged, 32-bit) with a cosine learning-rate schedule starting at 2e-4 and a 3\% warmup ratio. Effective batch size 16 (4 per device $\times$ 4 gradient accumulation steps). With \texttt{group\_by\_length=True} the run finished in roughly 40 minutes on a T4.

\textbf{Layer 2: Constrained decoding at inference time.} Even after fine-tuning, the model has the chatty habits of its instruction-tuned parent. To suppress them, we generate with very tight constraints: \texttt{max\_new\_tokens=20} (a hard cap on length, so the model literally cannot produce a paragraph), \texttt{temperature=0.1} (almost greedy decoding, no creative drift), and a repetition penalty of 1.1 (so the model does not loop on common tokens from the diff). This is the cheapest possible way to enforce conciseness, no extra weights, no extra compute, just stricter sampling.

\textbf{Layer 3: Deterministic post-processing.} Even with constrained decoding, the model sometimes leaks conversational filler (``Here is the commit:~...''). We apply three string-level fixes in sequence: tag slicing (keep only what comes after the final \texttt{[/INST]}), guillotine truncation (take the first line only, splits on the first newline), and conversational-prefix stripping (if there is a colon, look at the word before it; if that word is not in our valid Conventional Commits type set, strip everything up to and including that colon).

\subsection{Motivation for the Strategy}

Why this layered approach instead of just training harder?

\textbf{It is much cheaper.} Layers 2 and 3 cost nothing at training time and almost nothing at inference time, but they account for a large chunk of the final quality improvement. Our results (Section~\ref{sec:results}) show that the LLM-as-a-Judge score went from 1.97 (vanilla) to 2.20 (raw fine-tuned) to 3.68 (full pipeline). That is, fine-tuning on its own gave us a $+0.23$ improvement in judge score; the post-processing pipeline on top of the same weights gave us another $+1.48$. The pipeline is doing more work than the fine-tuning, for essentially zero additional compute.

\textbf{It is more controllable.} Deterministic post-processing is debuggable. If something goes wrong, we can read the log and find out which line of code did the wrong thing. If the same problem were caused by the model's weights, we would have to retrain. Pushing as much of the ``shape'' of the output into deterministic code as possible makes the system more predictable.

\textbf{It separates concerns.} The model's job is to understand the diff and produce a candidate message. The pipeline's job is to enforce the format. Separating these two jobs means we can swap out the model later (a Llama-3 7B, a Phi-3, a smaller distilled model) without rewriting the format-enforcement layer.

\textbf{Multi-metric evaluation was deliberate.} For the evaluation side, we used four different metrics, regex compliance, length, BERTScore, LLM-judge, because we did not trust any single metric to tell the truth. This decision turned out to be more important than we expected: our original compliance regex had two subtle bugs that flattered the baseline and penalized our pipeline (more on this in Section~\ref{sec:robust-metric}). If we had only reported compliance, we would have shipped a misleading result. Having BERTScore and LLM-judge as backup metrics meant we could cross-check.

% =====================================================================
\section{Implementation}

\subsection{Software Design}

The codebase is organized into seven logical phases:

\begin{enumerate}[leftmargin=1.5em,itemsep=0.2em]
  \item \textbf{Phase 0 -- Setup:} load the dataset and install dependencies.
  \item \textbf{Phase 1 -- Training:} load Mistral-7B-Instruct-v0.2 in 4-bit NF4, attach LoRA adapters, run \texttt{SFTTrainer} for one epoch, save the adapter.
  \item \textbf{Phase 2 -- Inference setup (clean session):} reload the base model + adapter on a fresh runtime to free training-time VRAM.
  \item \textbf{Phase 3 -- Batch generation:} generate predictions for the test slice using both the fine-tuned model and (separately) the vanilla base model with the LoRA adapter disabled.
  \item \textbf{Phase 4 -- Quantitative evaluation:} regex compliance, length distribution, BERTScore.
  \item \textbf{Phase 5 -- LLM-as-a-Judge:} async calls to Groq's Llama models with a 1--5 rubric.
  \item \textbf{Phase 6 -- Constrained decoding + post-processing:} run the same model with the aggressive inference parameters and apply the deterministic string-cleaning pipeline.
  \item \textbf{Phase 7 -- Final evaluation \& visualization:} re-run all metrics on the cleaned outputs and build the comparison plots.
\end{enumerate}

\begin{figure}[!t]
\centering
\includegraphics[width=\textwidth,height=0.28\textheight,keepaspectratio]{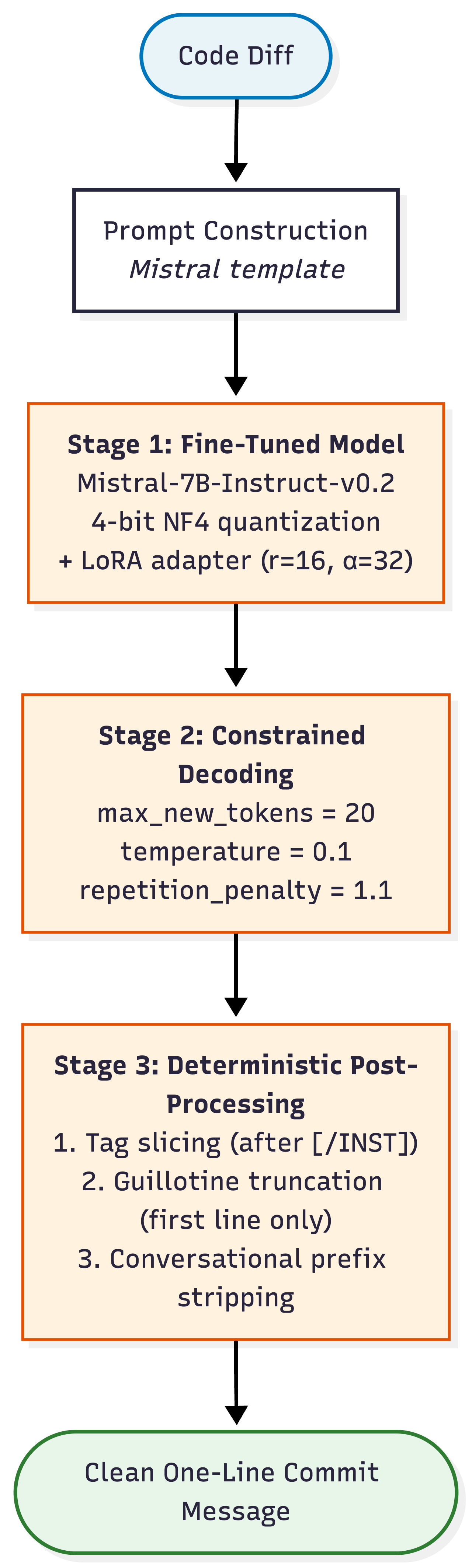}
\caption{End-to-end CommitLLM inference architecture. A code diff is wrapped in Mistral's chat template and passed through three stages: a LoRA fine-tuned base model, constrained decoding that caps output length and suppresses repetition, and deterministic string-level post-processing. Each stage adds \emph{constraints} rather than adding parameters, only Stage 1 involves learned weights.}
\label{fig:architecture}
\end{figure}

The post-processing function is short enough to include verbatim:

\begin{lstlisting}
VALID_TYPES = {"feat", "fix", "docs", "style", "refactor",
               "perf", "test", "chore", "build", "ci", "revert"}

def clean(raw_output):
    # Tag slicing
    s = raw_output.split("[/INST]")[-1] if "[/INST]" in raw_output else raw_output
    # Guillotine truncation
    s = s.strip().split('\n')[0]
    # Conversational prefix stripping
    if ":" in s:
        prefix = s.split(":", 1)[0]
        words = re.findall(r'[a-zA-Z]+', prefix)
        if words and words[-1].lower() not in VALID_TYPES:
            s = s.split(":", 1)[1]
    return s.strip()
\end{lstlisting}

\subsection{External Components}

We used the following third-party components:

\begin{itemize}[leftmargin=1.5em,itemsep=0.2em]
  \item \textbf{Mistral-7B-Instruct-v0.2}~\cite{jiang2023mistral} as the base model. Loaded from HuggingFace at runtime; we save and ship only the LoRA adapter delta.
  \item \textbf{CommitPackFT}~\cite{commitpackft} as the training/eval data. 9{,}600 train / 1{,}200 validation / 1{,}200 test examples, all in the \texttt{[INST] ... [/INST]} format already.
  \item \textbf{HuggingFace \texttt{transformers}, \texttt{peft}, \texttt{trl}, \texttt{datasets}, \texttt{evaluate}} for the modeling stack.
  \item \textbf{\texttt{bitsandbytes} ($\geq 0.46.1$)} for 4-bit NF4 quantization.
  \item \textbf{BERTScore} via the \texttt{evaluate} library, which uses RoBERTa-large under the hood as the embedding backbone.
  \item \textbf{Groq API} for LLM-as-a-Judge, calling \texttt{llama-3.3-70b-versatile} (raw outputs) and \texttt{llama-3.1-8b-instant} (clean outputs after hitting the daily token limit on the 70B model).
  \item \textbf{\texttt{pandas}, \texttt{matplotlib}, \texttt{seaborn}} for results aggregation and visualization.
\end{itemize}

\subsection{Experimental Design}

We compared three systems on the same 50-sample test slice (drawn from the first 50 examples of the CommitPackFT test split):

\begin{enumerate}[leftmargin=1.5em,itemsep=0.2em]
  \item \textbf{Vanilla Mistral} -- the base model with the LoRA adapter disabled. This is our baseline. It tells us how much fine-tuning matters.
  \item \textbf{CommitLLM (Raw Fine-Tuned)} -- the fine-tuned model with default decoding (\texttt{max\_new\_tokens=50}, \texttt{temperature=0.2}). This isolates the contribution of the LoRA training itself.
  \item \textbf{CommitLLM (Final Pipeline)} -- the fine-tuned model with constrained decoding ($\texttt{max\_new\_tokens=20}$, $\texttt{temperature=0.1}$, repetition penalty 1.1) plus the deterministic post-processing pass. This is the system as we would ship it.
\end{enumerate}

The evaluation was limited to 50 samples (out of 1{,}200 available in the test split) due to the Groq free-tier rate limit. Each LLM-as-a-Judge call sends the diff plus the ground truth plus the generated commit, which adds up to several thousand tokens per call. Across three systems and 50 samples, with the 2.1-second sleep between calls to stay under the per-minute rate, evaluation already takes about 6 minutes per pass. The deterministic metrics (compliance, length, BERTScore) would scale to the full 1{,}200 trivially; we kept the same 50 for consistency across all metrics.

We measured four things for every output:

\begin{enumerate}[leftmargin=1.5em,itemsep=0.2em]
  \item \textbf{Format compliance:} a regex-based rule for what a ``well-formed'' commit message looks like (Section~\ref{sec:robust-metric} for the corrected version).
  \item \textbf{Length:} character count, plus the fraction of outputs at or below 72 characters (the de-facto subject-line limit in git).
  \item \textbf{BERTScore (F1):} contextual-embedding similarity to the human-written ground-truth commit. Higher = closer to what a human would have written.
  \item \textbf{LLM-as-a-Judge score (1--5):} a strong LLM grades the generated commit against the diff and the ground truth, using a fixed rubric.
\end{enumerate}

% =====================================================================
\section{Results}
\label{sec:results}

\begin{figure*}[!t]
\centering
\includegraphics[width=.7\textwidth,height=0.32\textheight,keepaspectratio]{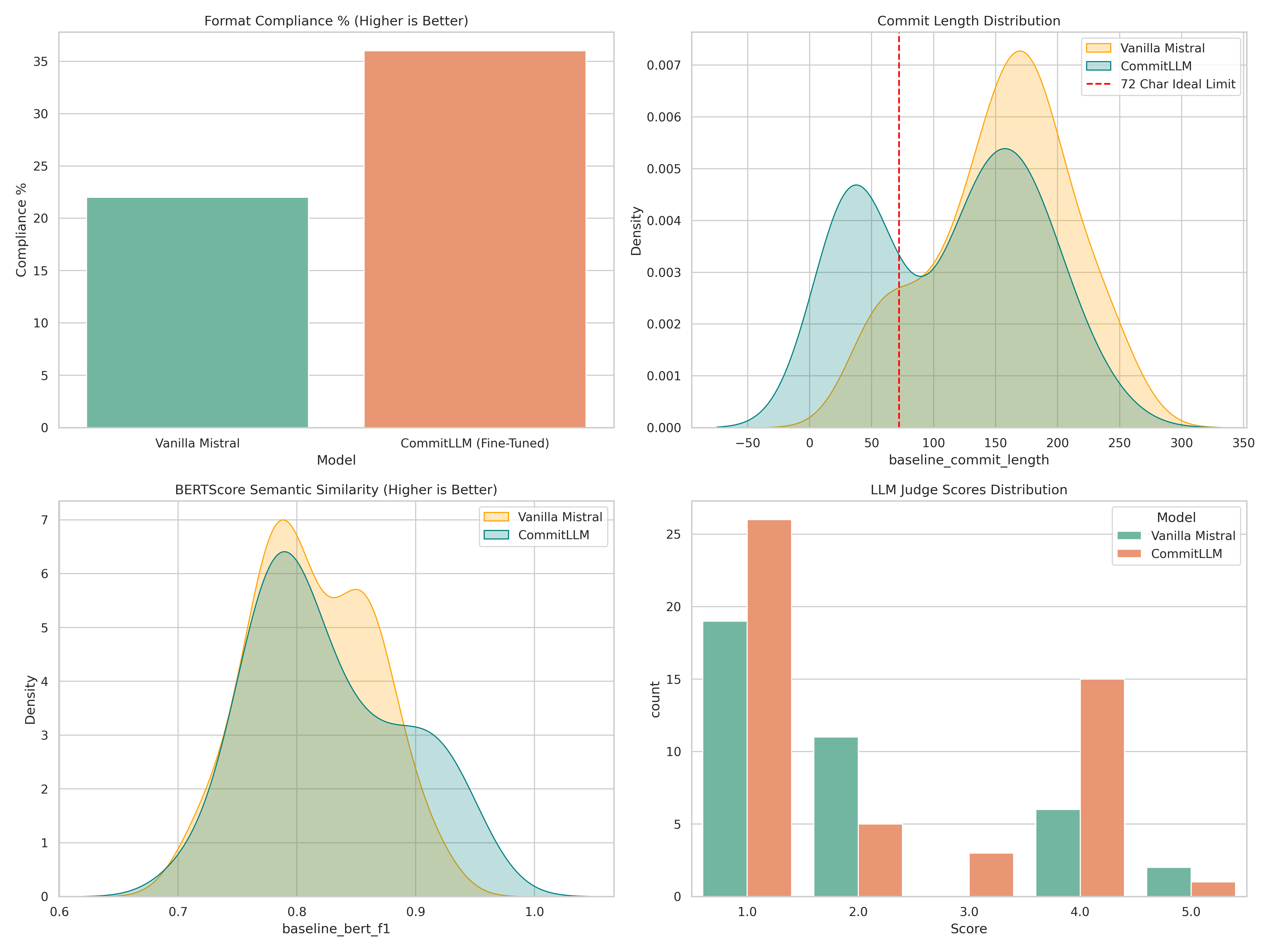}
\caption{Vanilla Mistral vs.\ raw fine-tuned model. Fine-tuning narrows the length distribution and shifts BERTScore upward, but the long tail of verbose outputs persists.}
\label{fig:baseline}
\end{figure*}

\begin{figure*}[!t]
\centering
\includegraphics[width=.7\textwidth,height=0.32\textheight,keepaspectratio]{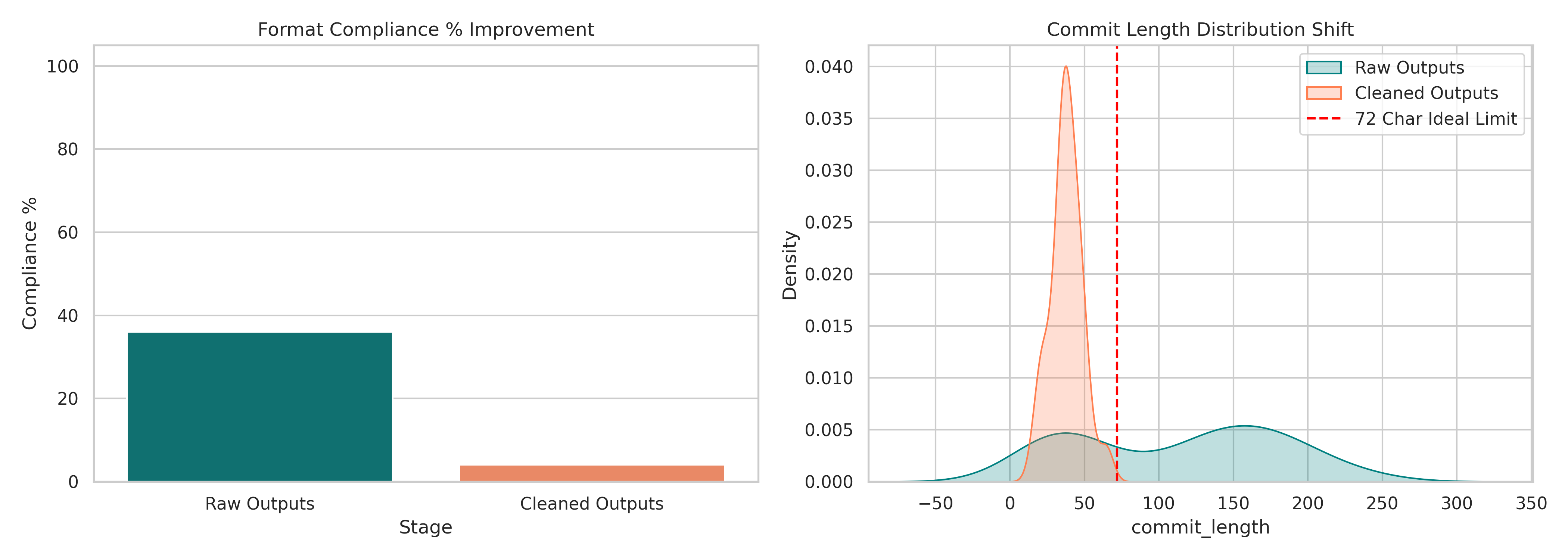}
\caption{Effect of the post-processing pipeline. Length distribution collapses onto a tight peak well below the 72-character ideal.}
\label{fig:post}
\end{figure*}

\begin{figure*}[!t]
\centering
\includegraphics[width=\textwidth,height=0.32\textheight,keepaspectratio]{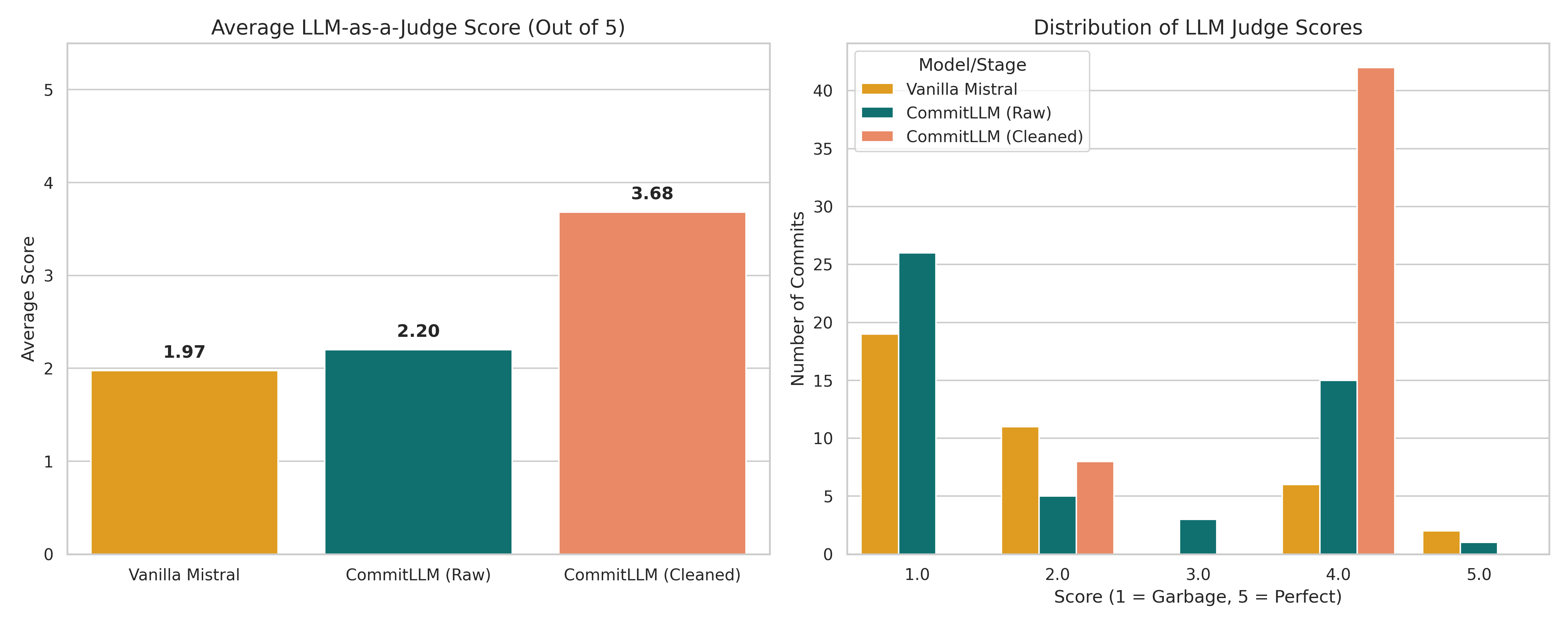}
\caption{Three-way LLM-as-a-Judge comparison. Score distribution shifts from the 1--2 range (``garbage / poor'') into the 3--5 range (``acceptable / perfect'').}
\label{fig:judge}
\end{figure*}

\begin{figure*}[!t]
\centering
\includegraphics[width=\textwidth,height=0.32\textheight,keepaspectratio]{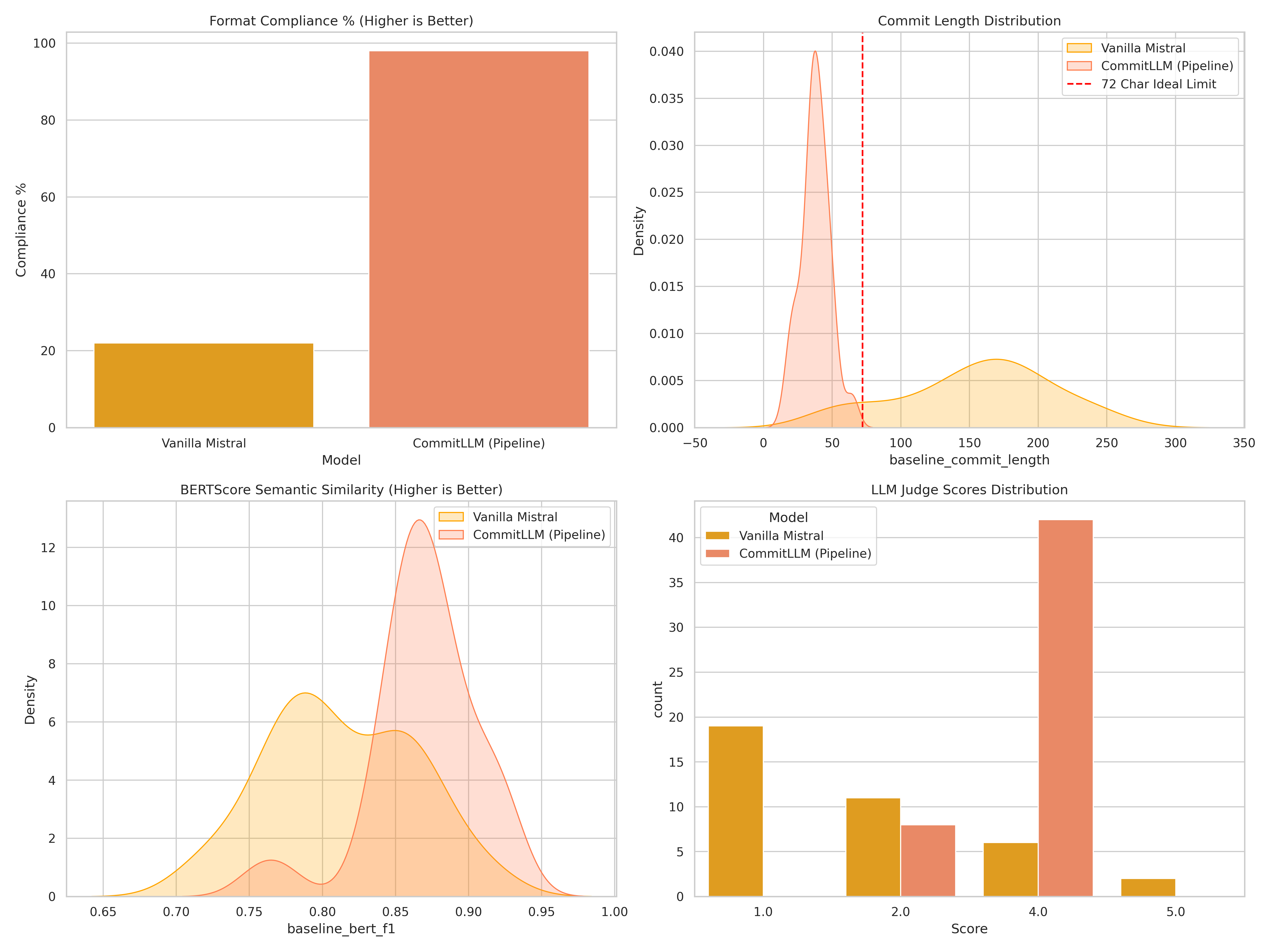}
\caption{Head-to-head: vanilla Mistral vs.\ the final CommitLLM pipeline on all four metrics.}
\label{fig:ultimate}
\end{figure*}

\subsection{Headline Numbers}

Table~\ref{tab:results} shows the main result. CommitLLM with the full pipeline beats the vanilla baseline on every metric, often by a large margin, and the post-processing layer accounts for most of the improvement.

\begin{table*}[h]
\centering
\caption{Comparison across three systems on the 50-sample test slice. ``Robust compliance'' uses the corrected metric from Section~\ref{sec:robust-metric}.}
\label{tab:results}
\renewcommand{\arraystretch}{1.15}
\begin{tabular}{lccc}
\toprule
\textbf{Metric} & \textbf{Vanilla Mistral} & \textbf{CommitLLM (Raw)} & \textbf{CommitLLM (Pipeline)} \\
\midrule
Format compliance (robust)  & 22.00\%       & 36.00\%       & \textbf{98.00\%} \\
Average length              & 154.8 chars   & 113.4 chars   & \textbf{37.9 chars} \\
\% under 72 chars           & $\sim$0\%     & 38.00\%       & \textbf{100.00\%} \\
BERTScore (F1)              & 0.8129        & 0.8259        & \textbf{0.8688} \\
LLM-Judge (avg / 5.0)       & 1.97          & 2.20          & \textbf{3.68} \\
\bottomrule
\end{tabular}
\end{table*}

A few observations from this table.

The fine-tuning by itself (vanilla $\rightarrow$ raw) is a modest improvement: BERTScore goes up by 0.013 and LLM-judge goes up by 0.23 points. The model has learned something about the structure of commit messages, but it is still verbose, average length only drops from 154.8 to 113.4 characters, which is still well past the 72-char ideal.

The pipeline on top of the same fine-tuned weights (raw $\rightarrow$ pipeline) is a much bigger jump: BERTScore goes up by another 0.043, LLM-judge by another 1.48 points. This is the clearest evidence that for structured output tasks, treating the LLM as a system with a deterministic post-processing layer pays off more than treating it purely as a model with weights.

\subsection{The Robust Compliance Metric}
\label{sec:robust-metric}

The format compliance number above (22\% / 36\% / 98\%) uses a corrected metric. Our original metric did not.

We started with a single regex:

\begin{lstlisting}
pattern = re.compile(
    r"^((feat|fix|docs|style|refactor|perf|test|chore|build|ci|revert)(\([^)]+\))?: .+)|(^[A-Z].+)$"
)
\end{lstlisting}

The first half of this regex (Conventional Commits prefixes) is fine. The second half, accept anything that starts with a capital letter, has two problems we did not catch until we looked at the data more carefully.

\textbf{False positives for the baseline.} Vanilla Mistral often produces multi-line paragraphs like \texttt{"Here is the commit message for your diff..."}. The regex sees \texttt{H} as the first character and rubber-stamps the whole paragraph as compliant, even though it is several lines of conversational text and would never pass a real commit hook. So the baseline's compliance number was artificially inflated.

\textbf{False negatives for the pipeline.} Our post-processing pipeline often produces commits like \texttt{"132: Add command line interface"} or \texttt{"2014-06-17: Fix bug"}, issue-ID-prefixed and date-prefixed commits that are extremely common in real-world repos and that appear in the training data. The original regex rejected these because they start with digits, not capital letters. So the pipeline's compliance number was artificially deflated.

The combined effect is the worst possible kind of measurement error: the metric was biased \emph{in favor of} the system we were comparing against and \emph{against} our own system. We replaced it with a stricter four-rule check (\texttt{is\_strictly\_compliant}):

\begin{enumerate}[leftmargin=1.5em,itemsep=0.2em]
  \item Reject any output containing a newline character (kills the multi-line false positives).
  \item Accept Conventional Commits with optional scope: \texttt{feat:}, \texttt{fix(ui):}, etc.
  \item Accept ID/date-prefixed commits: anything matching \texttt{[a-zA-Z0-9-]+: .+}.
  \item Accept concise single-line commits starting with a capital letter or digit.
\end{enumerate}

Under this corrected metric, the baseline correctly drops to 22\% (because most of vanilla Mistral's outputs were really paragraphs), and the pipeline correctly rises to 98\%. The model weights did not change between the two versions of the metric, only what we were asking the metric to measure changed. We think it is worth flagging this explicitly: a poorly written evaluation metric can hide a system's real performance, in either direction.

\subsection{The Long Tail of Verbose Outputs}

One observation that motivated the constrained decoding step: even after fine-tuning, the raw model's output length distribution still has a long tail past 200 characters. In other words, the average dropped, but a meaningful minority of outputs were still paragraph-length. For a production tool, the average is not what matters, the worst case is. A pre-commit hook that produces a 200-character paragraph one out of every four times is going to get disabled the same day it is installed.

The pipeline plot (post-processing comparison, Figure~\ref{fig:post}) shows what the constrained decoding plus post-processing does to that distribution: it collapses onto a tight peak in the 30--50 character range, with no tail past 72. This is exactly what we want for a commit-line generator.

\subsection{Why CommitLLM Wins on BERTScore}

BERTScore measures contextual-embedding similarity, so it should be tolerant of paraphrase, ``fix login bug'' and ``resolve authentication error'' would both score high against the same ground truth. So why does the vanilla baseline score lower (0.8129) than our pipeline (0.8688), if both models presumably understand the diff?

The answer is verbosity. Vanilla Mistral often buries the correct answer inside a wrapping paragraph (\textit{``Here is a commit message describing the change where...''}). Even when the right words appear, the surrounding filler dilutes the embedding similarity to the human reference. Our pipeline strips that filler, leaving a short string that aligns much more directly with the ground truth. So the BERTScore improvement is not (or not only) about the model getting better at understanding code, it is about the system getting better at concentrating the semantic signal into a commit-shaped string.

% =====================================================================
\section{Future Work}
\label{sec:future}

\textbf{IDE plugin.} The most natural deployment target is a VS Code extension that, on a button click in the source-control panel, reads the current staged diff, sends it through the pipeline, and pre-fills the commit message box. The most viable approach is to assume the user has a local inference server (e.g., Ollama or \texttt{llama.cpp}) running and have the extension talk to that endpoint.

\textbf{Full-test-set evaluation.} We evaluated on 50 samples instead of the full 1{,}200 in the test split because the LLM-as-a-Judge calls hit the free-tier daily token limit. With paid API access or a self-hosted judge model, scaling to the full test set would be straightforward.

\textbf{More training epochs.} We trained for one epoch only. Loss curves suggested the model was still improving at the end of training. A 2--3 epoch run with validation-loss-based early stopping might give a meaningfully better raw fine-tuned model and reduce how much work the post-processing layer has to do.

\textbf{GGUF conversion for local deployment.} Once the LoRA adapter exists, the standard recipe for local inference is to merge the adapter into the base, quantize the merged model to GGUF format, and run it under \texttt{llama.cpp}. This is a prerequisite for the IDE plugin path described above.

\textbf{Human evaluation.} The LLM-as-a-Judge scores are a proxy. The true north-star metric is whether actual developers prefer CommitLLM's output to either their own or to the vanilla baseline. A user study with developers blind-rating commits on a representative diff set would make the results more robust.

\textbf{Hallucination filtering for fake issue IDs.} Because CommitPackFT contains real commits like \texttt{"132: Add ..."}, our model sometimes emits commits with plausible but fabricated issue IDs. A regex filter in post-processing could strip these, or alternatively a learned classifier trained on a small set of labeled examples.

% =====================================================================
\section{Conclusion}

We presented CommitLLM, a three-stage pipeline for generating git commit messages from code diffs. By combining QLoRA fine-tuning of Mistral-7B with constrained decoding and deterministic post-processing, the system achieves 98\% format compliance, an average output length of 37.9 characters, and an LLM-as-a-Judge score of 3.68/5---substantially outperforming the vanilla baseline on every metric. The key finding is that the deterministic pipeline layers (constrained decoding + post-processing) contribute more to output quality than the fine-tuning itself: the judge score improved by $+0.23$ from fine-tuning alone, but by $+1.48$ from the pipeline on top. This suggests that for structured-output tasks like commit message generation, engineering the system around the model matters more than engineering the model alone. The entire pipeline runs on a single NVIDIA T4 GPU, making it accessible to individual developers without cloud API dependencies.

% =====================================================================
\bibliographystyle{plain}

\end{document}